\documentclass[a4paper,11pt]{article}
\pdfoutput=1 
\usepackage{jinstpub} 
\usepackage{lineno}
\usepackage{url}
\usepackage{graphicx}
\usepackage{caption}
\usepackage{float}
\usepackage{subfig}
\usepackage{appendix}
\usepackage{hyperref}
\usepackage{xspace} 

\title{\boldmath The cryogenic RWELL: a stable charge multiplier for dual-phase liquid argon detectors}

\author[a,1]{A. Tesi\note{Corresponding author}}
\author[a,b]{, S. Leardini}
\author[a]{, L. Moleri}
\author[b]{, D. Gonzalez-Diaz}
\author[a]{, A. Jash}
\author[a]{, A. Breskin}
\author[a]{and S. Bressler}

\affiliation[a]{Department of Particle Physics and Astrophysics, Weizmann Institute of Science, Rehovot, Israel}
\affiliation[b]{Instituto Galego de Física de Altas Enerxías, Univ. de Santiago de Compostela, Santiago de Compostela, Spain}

\emailAdd{andrea.tesi@weizmann.ac.il}

\abstract{The operation of a cryogenic Resistive WELL (RWELL) in liquid argon vapor is reported for the first time. It comprises a Thick Gas Electron Multiplier (THGEM) structure coupled to a resistive Diamond-Like Carbon (DLC) anode deposited on an insulating substrate. The multiplier was operated at cryogenic temperature (90~K, 1.2~bar) in saturated argon vapor and characterized in terms of charge gain and electrical stability. A comparative study with standard, non-resistive THGEM (a.k.a LEM) and WELL multipliers, confirmed the RWELL advantages in terms of discharge quenching - i.e. superior gain and stability.}


\begin{document}
\maketitle
\flushbottom

\section{Introduction}
\label{sec:intro}
In recent years, the Thick Gaseous Electron Multiplier (THGEM) \cite{Breskin_2010, Breskin_2009} and its derivatives have become a leading technology in the field of particle and radiation detection. Thanks to their robustness and relatively low cost, these detectors have been proposed and are currently employed in numerous fields, as reviewed in \cite{Bressler_2023}. Amongst its derivatives, several attempts were conducted to couple the THGEM to resistive materials in order to enhance the detector's immunity against electrical discharges. Closed, WELL-like configurations \cite{Bellazzini} were proposed with a single-sided THGEM structure attached to a resistive anode, decoupling the multiplier from the readout electrode. Amongst them, the Resistive WELL (RWELL) \cite{Arazi_2014} can be found, having a thin resistive layer anode decoupled from the readout by an insulator; another one is the Resistive Plate WELL (RPWELL) \cite{Rubin_2013}, with the resistive-plate anode directly in contact with the readout electrode. Others include the Segmented RWELL (SRWELL) \cite{Arazi_2013_segmented}, a variant of the RWELL featuring a metallic grid under the resistive film (for faster charge evacuation), that, together with the RPWELL, were proposed as potential sampling elements for Digital Hadronic Calorimetry (DHCAL) \cite{Arazi_2012, Shikma2013, Bressler_2020}.
All these THGEM and THGEM-like concepts were extensively investigated at room temperature.  THGEM detectors (LEM, Large Electron Multipliers) were investigated very intensively in dual-phase liquid argon TPCs \cite{LEM_1}, as potential candidates for the DUNE neutrino experiment \cite{Falcone}. Results of extensive systematic studies of THGEM-based detectors in noble-liquid vapors are reported in \cite{Buzulutskov_2012, Buzulutskov_2020}. Others were studied at liquid-xenon temperatures in counting gases, e.g. Ne/CH$_4$; among them in Gas Photo-Multipliers (GPM) \cite{Arazi_2015_Xe, Israelashvili_2017}. The RPWELL was successfully investigated in Ne/CH$_4$ down to 163K \cite{Roy2019} and resistive materials have been explored also for large-volume experiments as a possible solution to electrical discharges \cite{RD51_2020, RD51_2020_Rui}.\\

\noindent
 Notwithstanding, the charge readout in dual-phase detectors has been practically abandoned, at least in pure Ar vapors, due to instabilities even at very low gains \cite{Autiero_DUNE}. A successful concept for attaining robust operation at higher charge gains could permit, e.g. in neutrino experiments, lowering detection thresholds thus reaching sensitivities superior to those of current single-phase devices \cite{Kim}. A small increase of the charge gain, though, would likely remain insufficient for dark-matter experiments, dealing with very low WIMP-deposited energies \cite{Aalbers_DM}. The possibility of integrating resistive materials into detector assemblies, to mitigate the disruptive effects of sparks, turned out to be less straightforward than expected - due to the insulating behavior of resistive materials at low temperatures. So far, only the operation of an RPWELL equipped with a Fe$_2$O$_3$/YSZ ceramic plate was reported in Ne/CH$_4$ at liquid xenon temperature (163~K) \cite{Roy2019}. A game-changing solution that made possible the operation of resistive detectors at liquid argon temperature (90~K) was provided by Diamond-Like-Carbon (DLC) layers \cite{ROBERTSON}. This material has been employed lately for spark protection in various other gas-avalanche detectors at room temperature \cite{Bencivenni_2023, Bencivenni_2019, Iodice_1, Maggi_22, Colaleo_19}.  
By adjusting the thickness of the DLC layer, it was possible to cover a broad range of resistivities suitable for spark protection, down to the liquid-vapor coexistence point of argon (87.5~K at 1~bar). 
In this work, the results of the successful operation of a cryogenic RWELL are summarized and an overview of the detector properties in comparison to its non-protected counterparts (THGEM, THWELL) is provided.

\section{Experimental setup}

\subsection{Detector assembly and readout}
\begin{figure}[H]
    \centering
    \includegraphics[width=15cm]{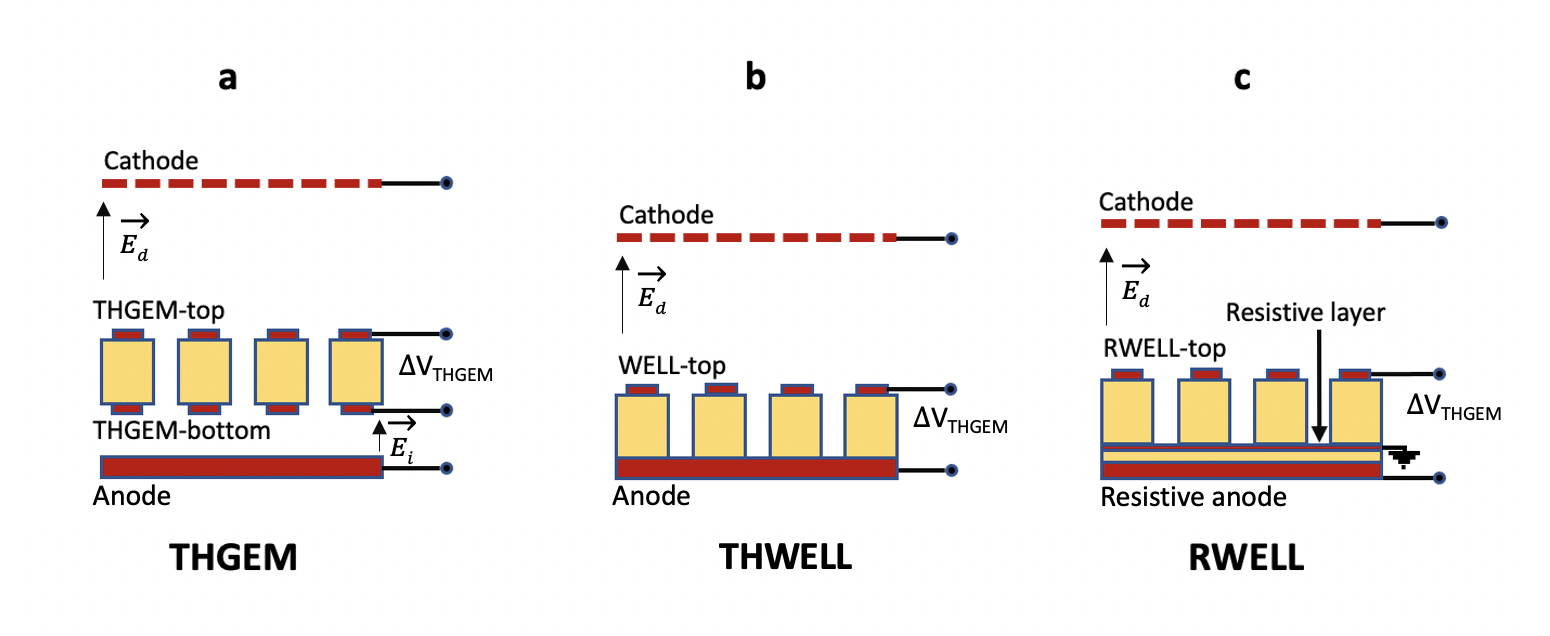}
    \centering
    \caption{\footnotesize Scheme of {\bf{(a)}} a THGEM (LEM) detector followed by an induction gap and a readout anode; {\bf{(b)}} a THWELL detector, with a metallic anode at its bottom; {\bf{(c)}} a RWELL detector, with a thin-film resistive anode decoupled from the metallic anode by an insulating plate. In all three cases, the drift gap was equal to 15~mm and the THGEM structure (double-sided in a) and single-sided in b), c)) was 0.8~mm thick.} 
    \label{fig:Structures}
\end{figure}

During this study, three detector concepts were investigated: THGEM, THWELL, and RWELL; they are schematically presented in Fig.~\ref{fig:Structures}.

\noindent
The first investigated configuration consisted of a double-sided 3x3~cm$^2$, 0.8~mm thick THGEM (0.5~mm diameter holes distributed in a hexagonal pattern with 1~mm pitch, and $\sim$0.1~mm hole rim) followed by an induction gap of 2~mm and an induction field E$_i$~=~5~kV/cm (see Fig. \ref{fig:Structures}a). In all the cases, the drift field was E$_d$~=~0.5~kV/cm.
The second structure was a 0.8~mm thick THWELL detector: a 3x3~cm$^2$, 0.8~mm thick single-sided THGEM structure (0.5~mm diameter holes distributed in a hexagonal pattern with 1~mm pitch and $\sim$0.1~mm hole rim) directly coupled to a metallic anode (see Fig.~\ref{fig:Structures}b).
The RWELL was realized with a 3x3~cm$^2$, 0.8~mm thick single-sided THGEM electrode (with the same geometry of the THWELL) pressed onto a metallic anode via a resistive layer deposited on a 0.1~mm thick Kapton insulating foil (see Fig.~\ref{fig:Structures}c).
Experiments were conducted using DLC films with different surface resistivities R$_S$ at room temperature, "RT" (i.e., from choosing different thicknesses during sputtering) in order to assess the quenching properties at cryogenic temperatures. More details about the DLC production process can be found in \cite{SONG2020}. These layers exhibited a stretched-exponential behavior as a function of temperature, with a coefficient a=1/3 (attributable to two-dimensional variable-range electron hopping).
Additional details regarding the properties of the DLC coatings at cryogenic temperature can be found in \cite{Leardini2022}. The responses of RWELL configurations with lower resistivity (e.g., $\mathrm{R_S^{DLC}}\sim$165~k$\Omega/\square$ at 298~K and $\mathrm{R_S^{DLC}}~\sim$10~M$\Omega/\square$ at 90~K, with $\square$ = 1 cm$^2$) were found to be similar to those of non-resistive readouts and thus are not reported here. 
In Table~\ref{table_RL}, the values for the relevant R$_S$ at 298~K and at 90~K are given. 

\begin{table}[H]
\begin{center}
\begin{tabular}{ |c|c|c| } 
\hline
DLC layer &  $\mathrm{R_S^{DLC}}$  at 298~K & $\mathrm{R_S^{DLC}}$  at 90~K \\
\hline
Sample 1 & $\sim14$~M$ \Omega /\square$ & $\sim20$~G$ \Omega /\square$ \\ 
Sample 2 & $\sim1$~M$ \Omega /\square$ & $\sim200$~M$ \Omega /\square$ \\ 
%Sample 3 & $\sim40$M$ \Omega /\square$ & $\sim200$G$ \Omega /\square$ \\
\hline
\end{tabular}
\end{center}
\caption{\footnotesize Surface resistivity at 298~K and at 90~K for two different DLC samples.   }
\label{table_RL}
\end{table}

\begin{figure}[H]
    \centering
    \includegraphics[height=6cm,width=7cm]{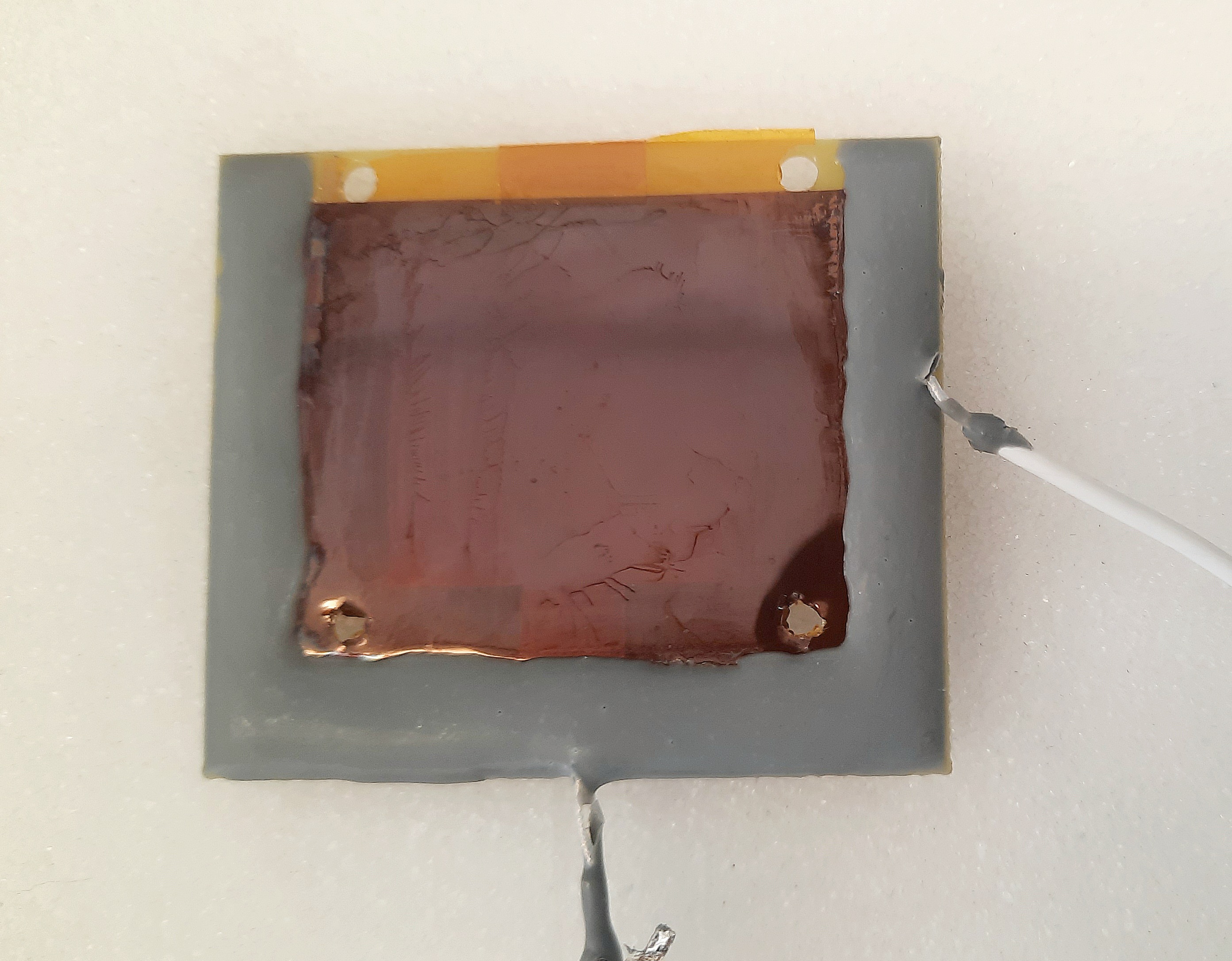}
    \centering
    \caption{\footnotesize Resistive anode composed by a DLC layer fixed on a PCB board with electrically-insulating cryogenic epoxy. The electric connections, realized with cryogenic conductive epoxy, are encapsulated and connected with coaxial wires.} 
    \label{fig:anode_RWELL}
\end{figure}

\noindent
To create electrical contacts, the DLC-coated Kapton foil was fixed onto a PCB, and the DLC layer was connected to two sides of copper lines using cryogenic conductive epoxy\footnote{Master Bond EP21TDCS-LO}. An electrically-insulating cryogenic epoxy\footnote{Stycast 2850FT}, was subsequently used to encapsulate the components (see Fig.~\ref{fig:anode_RWELL}). \\

\noindent
All the electrodes (i.e. cathode, THWELL, resistive layer and readout electrode) were biased by an HV power supply\footnote{CAEN N1471H} or via a charge-sensitive preamplifier\footnote{Cremat: Model CR-110 with CR-150-R5 evaluation board}. 
The latter was equipped with a protection circuit composed of surge arresters\footnote{8EC 90, EPCOS-TDK Electronics} in parallel and a series resistor of 220~$\Omega$. 
The time evolution of the detector gain was studied in order to establish stable operation conditions. For gain stabilization studies, the signals from the preamplifier were fed to an amplifier\footnote{Ortec Model 450}; the amplified signals were digitized by a multi-channel analyzer\footnote{Amptek MCA 8000D}. For all the other measurements, waveforms from the preamplifier were acquired using a digital oscilloscope\footnote{Tektronix MSO 5204B} and post-processed using dedicated Matlab scripts. Histograms of the preamplifier signal amplitude were also acquired using the oscilloscope. 
Currents from the electrodes were read out by a digitizer\footnote{NI USB-6008} and monitored using LabVIEW SignalExpress \cite{Labview}.

\subsection{Cryostat}
\label{subparagraph: Measurements methodology}
\begin{figure}[H]
    \centering    \includegraphics[height=11cm,width=12cm]{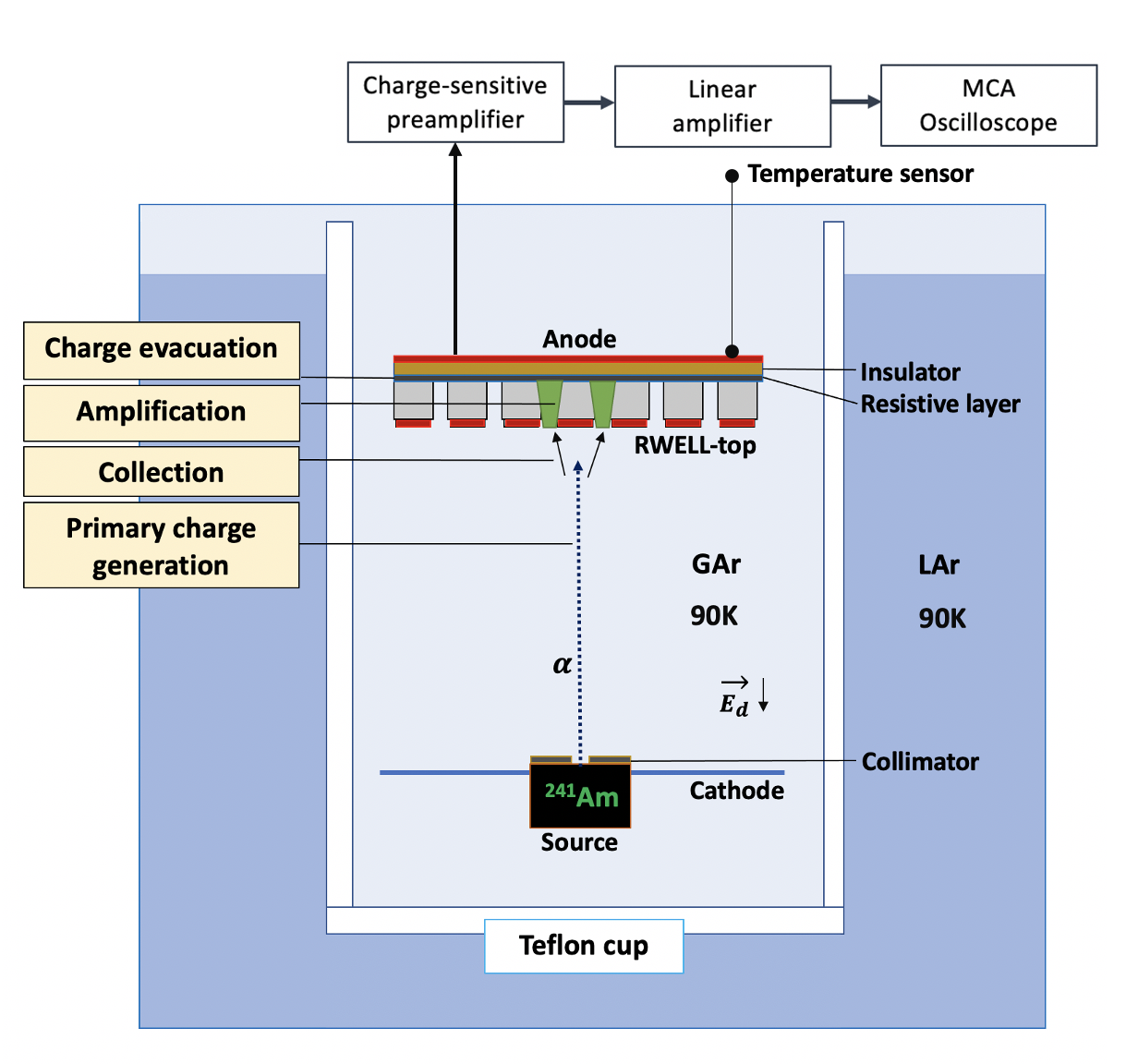}
    \centering
    \caption{\footnotesize Scheme of the cryogenic setup for investigating the detectors (here an RWELL). The assembly and the involved physical processes are detailed. The detector was operated at 90K, 1.2~bar.} 
    \label{fig:Setup_RWELL}
\end{figure}

\noindent
The measurements were performed in a dedicated LAr cryostat, WISArD (Weizmann Institute of Science liquid Argon Detector) described in detail in \cite{Erdal_2019}. In all the experiments, an $^{241}$Am source emitting 5.5~MeV alpha particles with a rate of about 10~Hz was installed on a metallic plate cathode at a distance of 15~mm from the multiplying electrode. The source emission was opportunely confined using a metallic collimator of 4~mm thickness and 5~mm diameter. A 12~$\mathrm{\mu m}$ Mylar foil was installed at the collimator's emitting face to attenuate the alpha energy to about 4~MeV. This granted full containment of the alpha-particles in the drift volume ($\approx$7 mm range in gaseous argon at 90~K, 1.2~bar calculated with SRIM \cite{SRIM} assuming a gas density of 6.65 kg/m$^3$ \cite{NIST}. Uncertainties related to the longitudinal and lateral straggling accounted for $\approx$280$\mu$m and $\approx$300$\mu$m, respectively.). 
Each detector configuration was operated in the saturated vapor phase of argon at 90~K. This was realized by inserting the detector assembly inside a Teflon cup, as depicted in Fig.~\ref{fig:Setup_RWELL}. Argon was liquefied into the cryostat using a cryocooler\footnote{Cryomec PT90}, till the upper edge of the cup. The level of the liquid was monitored using temperature sensors located in the proximity of the cup edges, inside the cup adjacent to the detector anode, and in other parts of the system. This prevented the liquid from penetrating into the detector region. The chosen operation within the vapor enclosed in the cup rather than in the liquid phase minimized losses due to electron recombination and extraction (see Fig.~3 in \cite{Aprile_2010}), thus yielding a higher number of ionization electrons and facilitating the measurements.
The temperature inside the system was constantly monitored using a temperature controller\footnote{CryoCon model 24}. A conceptual scheme of the cryogenic setup is depicted in Fig.~\ref{fig:Setup_RWELL}.

\section{Methodology}
\subsection{Physics of the detector}
The 4~MeV alpha particles (attenuated $^{241}$Am source) generate $\sim$10$^5$ primary electrons in the drift region along a track forming an angle of at least 60$^\circ$ with the detector surface (W-value of 26.3~eV in Ar \cite{Doke}).
These are drifted towards the THGEM holes under a field $\mathrm{E_d = 0.5}$~kV/cm. Increasing the bias across the multiplier electrode, defined as $\mathrm{\Delta V_{THGEM}}$ results in a high electric field within the holes, allowing for the onset of avalanche multiplication of the primary charges. The amplification factor (i.e. detector gain) increases exponentially with $\mathrm{\Delta V_{THGEM}}$, typically in the range of 2.7-3.2~kV in our configuration. Charge from avalanche electrons is evacuated to the ground through the anode in the THGEM and THWELL multipliers and along the resistive-anode surface in the RWELL (see Fig.~\ref{fig:Structures}). An electrical signal is induced onto all the electrodes by the movement of charges (electrons and ions) during the entire process, as described by the Shockley-Ramo theorem \cite{Riegler_2016}. 
We refer to the collection signal as the one induced by the movement of the primary charges in the drift gap. This was recorded from the cathode in specific measurements where $\mathrm{\Delta V_{THGEM}}$ = 0~kV. We refer to the amplification signal as the one induced by the movement of the avalanche charges. It was recorded from the anode with $\mathrm{\Delta V_{THGEM}}$ in the range of detector operation. The effective detector gain can be estimated by the following normalization:
\begin{equation}
  \mathrm{  G_{Eff} = \frac{P_{Amplif}}{P_{Coll}}
  } 
  \label{eqn:gain}
\end{equation}
\noindent
where P$\mathrm{_{Amplif}}$ and P$\mathrm{_{Coll}}$ are the Gaussian mean values of the spectra of amplification and collection signals, respectively.
Notice that the anode signal of a THGEM operated in multiplication mode is mostly due to the fast movement of electrons within the induction gap - typically 4~$\mu$s in our setup at cryogenic temperature - whereas in THWELL and RWELL the signals contain both electron and slower ion components \cite{Bressler_2023}.
It is important to mention that the resistive anode, decoupled from the readout electrode by a thin insulator (here 0.1~mm thick Kapton), causes a small decrease in the effective gain because of a reduced weighting potential. This effect is negligible when the insulator is thin \cite{Jash:2022bxy}.

\subsection{Gas purity}
Gas purity plays a crucial role in LAr-TPCs, where a sub-ppb level of electronegative impurities is targeted \cite{Majumdar_2021} in order to minimize electron capture during the drift in the liquid or gas. The maximum achievable gain of gaseous multipliers in Ar is strongly affected by the presence of impurities, quenching for instance avalanche-induced VUV photons that are responsible for secondary avalanche feedback \cite{Miyamoto_2010}. Therefore, for a realistic comparison of the multipliers, they should be investigated in very pure Ar vapor. For this reason, prior to any measurement, the system was vacuum-pumped for at least 12~h reaching a pressure of $\sim$1$\times$10$^{-4}$~mbar.
A residual gas analyzer\footnote{SRS RGA200} was used to measure the level and composition of residual impurities. It was found that the main sources of contamination were H$_2$O, Ne, HF, N$_2$, CO. When filling the system with $\sim$1 bar of Ar, the residual impurities got diluted to a concentration of 0.1~ppm. This should be compared with the level of impurities in the Ar gas bottle: 99.999$\%$ with H$_2$O $\le$ 2~ppm, O$_2$ $\le$ 2~ppm, C$_n$H$_m$ $\le$ 0.5~ppm and N$_2$ $\le$ 5~ppm, as specified by the producer. During operation, the gas was recirculated and purified with a hot getter\footnote{Entegris HotGetter PS3-MT3-R-2} in order to grant a nominal impurity content of the order of 1~ppb.\\

\subsection{Liquefaction and thermal stabilization}
\label{liquefaction}
The cryostat was filled up with liquid argon until a temperature of 90 K was measured in the gaseous phase at the level of the assembly (temperature sensor in Fig.~\ref{fig:Setup_RWELL}). The pressure inside the cryostat throughout all the operations corresponded to the saturation pressure P$_s$ = 1.2~bar at 90~K. 
In order to grant thermal stability, according to our experience, the detector was not operated during the first six hours following liquefaction. During this period, the gas was purified at 2~L per hour for at least 24 hours.\\

\subsection{Gain stabilization and voltage scan}
\begin{figure}[H]
    \centering
    \includegraphics[width=15cm]{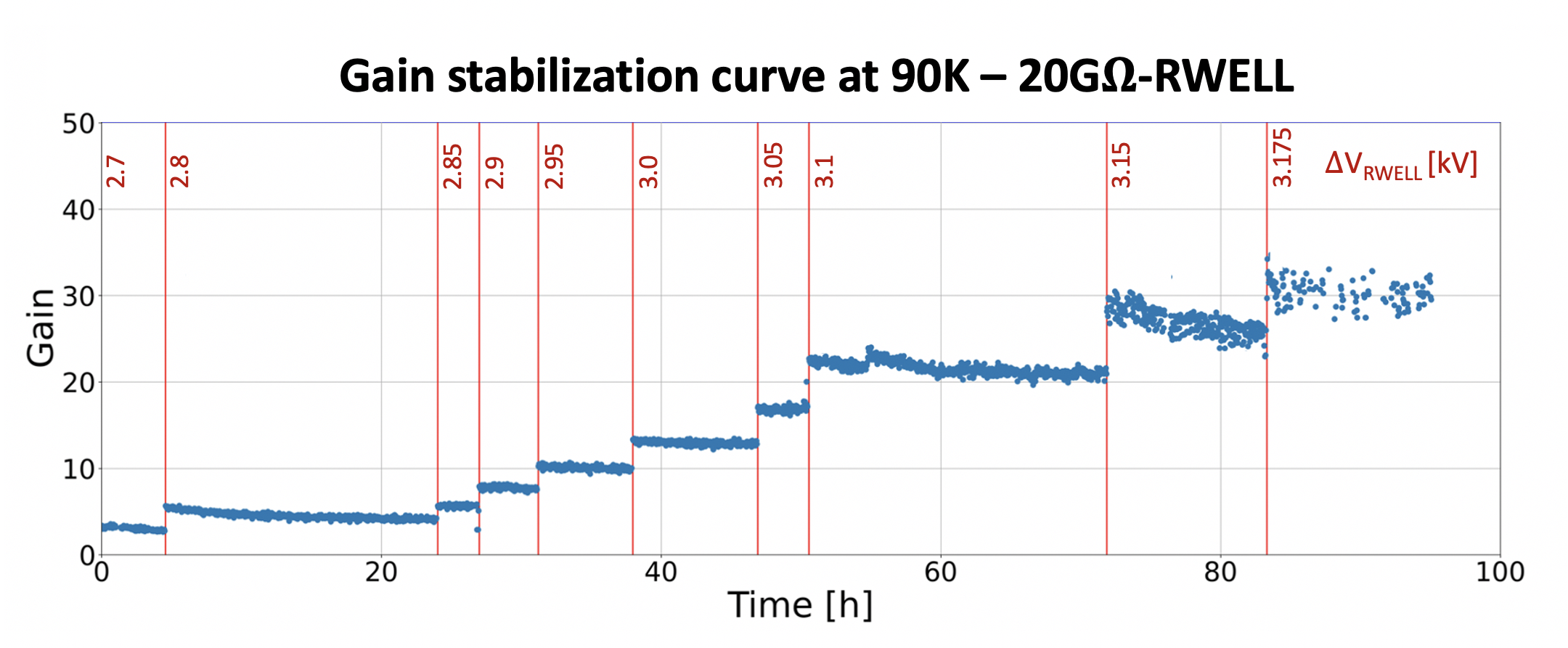}
    \centering
    \caption{\footnotesize Gain stabilization for the 20G$\Omega$-RWELL detector operated with alpha-particles at 90~K, 1.2~bar. The voltage configurations are indicated by red vertical bars and the corresponding values in kV are reported above them. The gain decreased over time (20$\%$ gain reduction) until stabilization. For 3~kV < $\mathrm{\Delta V_{RWELL}}$ < 3.175~kV, the detector operated in the presence of quenched discharges. Gain fluctuations are also visible at $\mathrm{\Delta V_{RWELL}}$~=~3.15 and 3.175~kV. }
    \label{fig:GainStab}
\end{figure}

The gain of each detector configuration was characterized by scanning the voltage across the RWELL, $\mathrm{\Delta V_{RWELL}}$, in the operational range while E$_d$ = 0.5~kV/cm was kept constant.
An increase of $\mathrm{\Delta V_{RWELL}}$ corresponded to an increase of the detector gain followed by a small reduction (about 20$\%$), up to stabilization after several hours. This effect is due to the accumulation of charges on the THGEM's insulating substrate (holes' walls and rims) \cite{Renous_2017}. 
For each voltage configuration, a minimum of 100 spectra of 120~s each were acquired. In Fig.~\ref{fig:GainStab}, the gain stabilization for the RWELL detector operated at 90~K, 1.2~bar is shown as an example. $\mathrm{\Delta V_{RWELL}}$ was set to 2.7~kV, sufficient to obtain well-defined spectra above noise, and it was subsequently increased up to 3.175~kV in steps of 50~V or 100~V. 
After a certain value of $\mathrm{\Delta V_{RWELL}}$, the detector entered a region of electrical instability. This was different for each detector configuration. For the THGEM and THWELL, a single discharge made the detector unstable by removing the charging-up, and a power cycle was required.%Moreover, the discharge currents were large enough to inhibit the power supply operation until restarted. 
For the RWELL, discharges were quenched by the resistive layer and the detector could be operated stably despite a loss in energy resolution due to small gain fluctuations (e.g. $\mathrm{\Delta V_{RWELL}}$ = 3.15 and 3.175~kV in Fig.~\ref{fig:GainStab}). 

\noindent
For the RWELL, three regions of operation were identified: i) discharge-free behavior up to 3~kV, ii) stable operation in the presence of quenched discharges for 3~kV < $\mathrm{\Delta V_{RWELL}}$ < 3.175~kV and iii) presence of constant currents above 3.175~kV. The measurement was stopped when the detector entered the latter region. For the THGEM and THWELL, two regions of operation were identified: discharge-free behavior up to $\mathrm{\Delta V_{THGEM}}$ = 3.2~kV and $\mathrm{\Delta V_{THWELL}}$ = 2.9~kV, and an unstable region of intense discharges above it; there the measurements were stopped.

\section{Results}
\subsection{Collection signals, spectra, and efficiency}
\begin{figure}[H]
    \subfloat{
       \includegraphics[width= 7.5 cm]{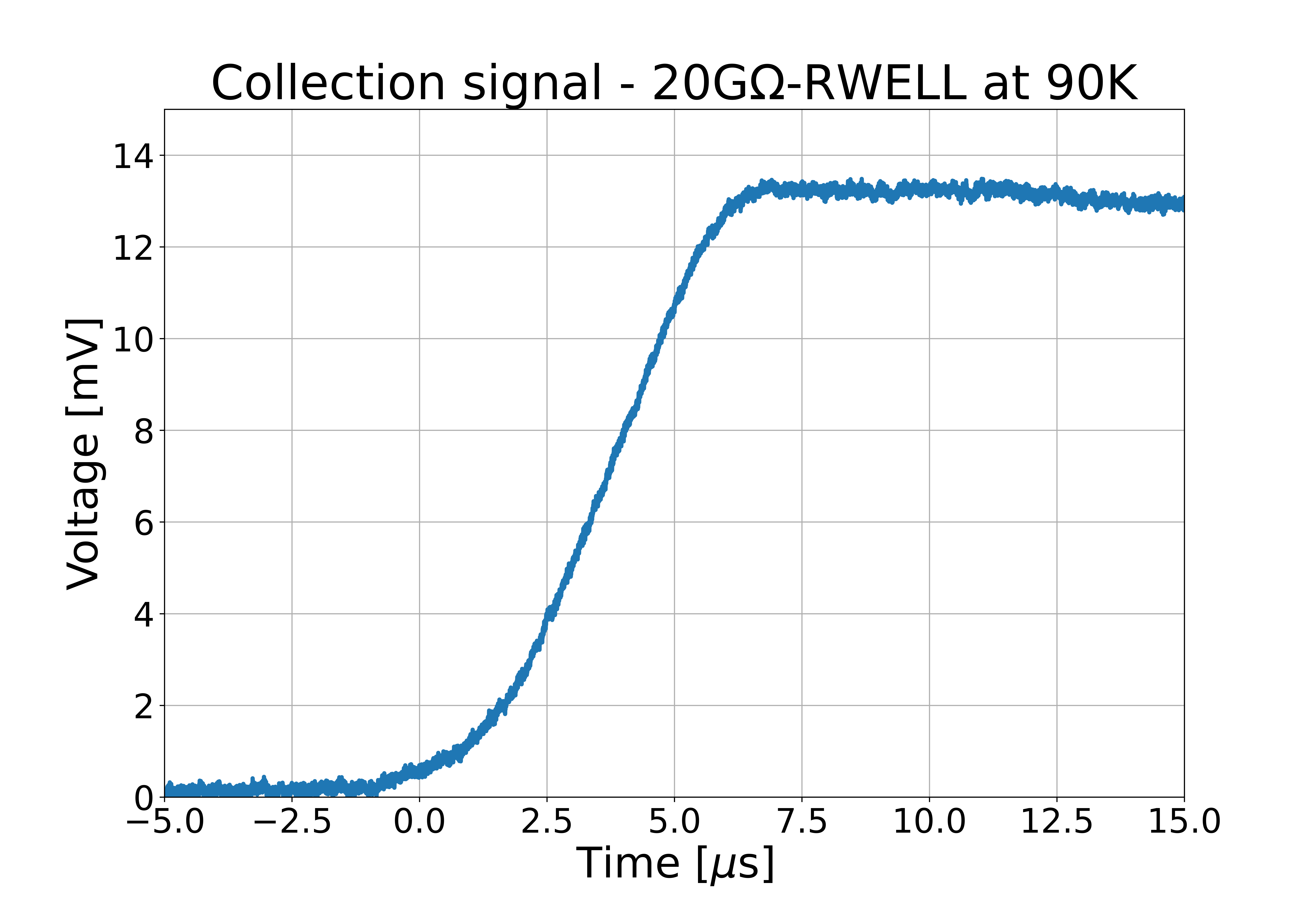}}
       \hfill
      \subfloat{
    \includegraphics[width=7.5 cm] {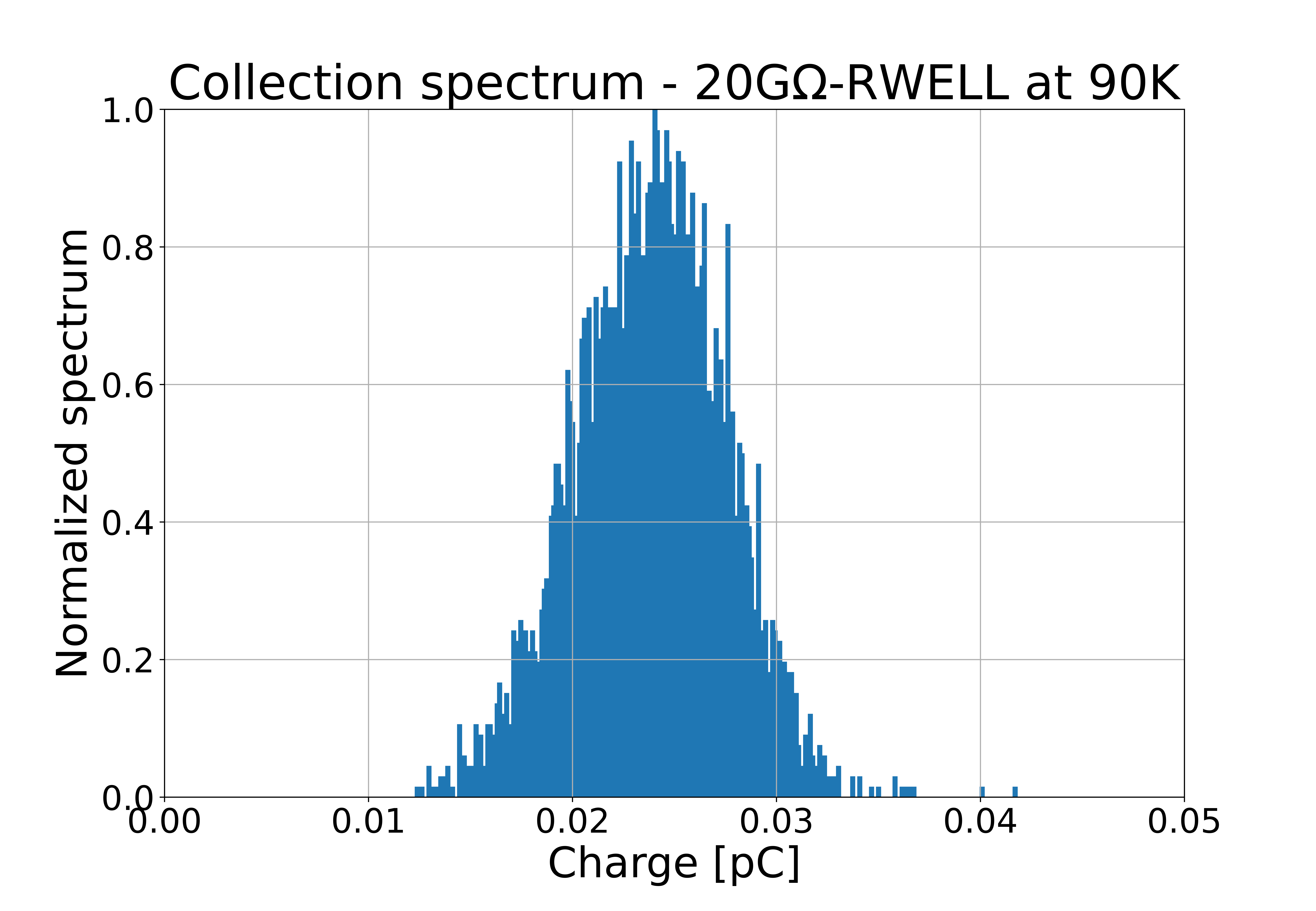}}
    \caption{\footnotesize \textbf{Left}: Average charge-collection waveform from the preamplifier, for an RWELL detector operated with alpha-pathe spectrumicles at 90~K, 1.2~bar; \textbf{Right}: signal amplitude spectrum calibrated to charge.} 
    \label{fig:CollectionSignals}
\end{figure}

\begin{figure}[H]
\subfloat{
\includegraphics[width=7.5cm]{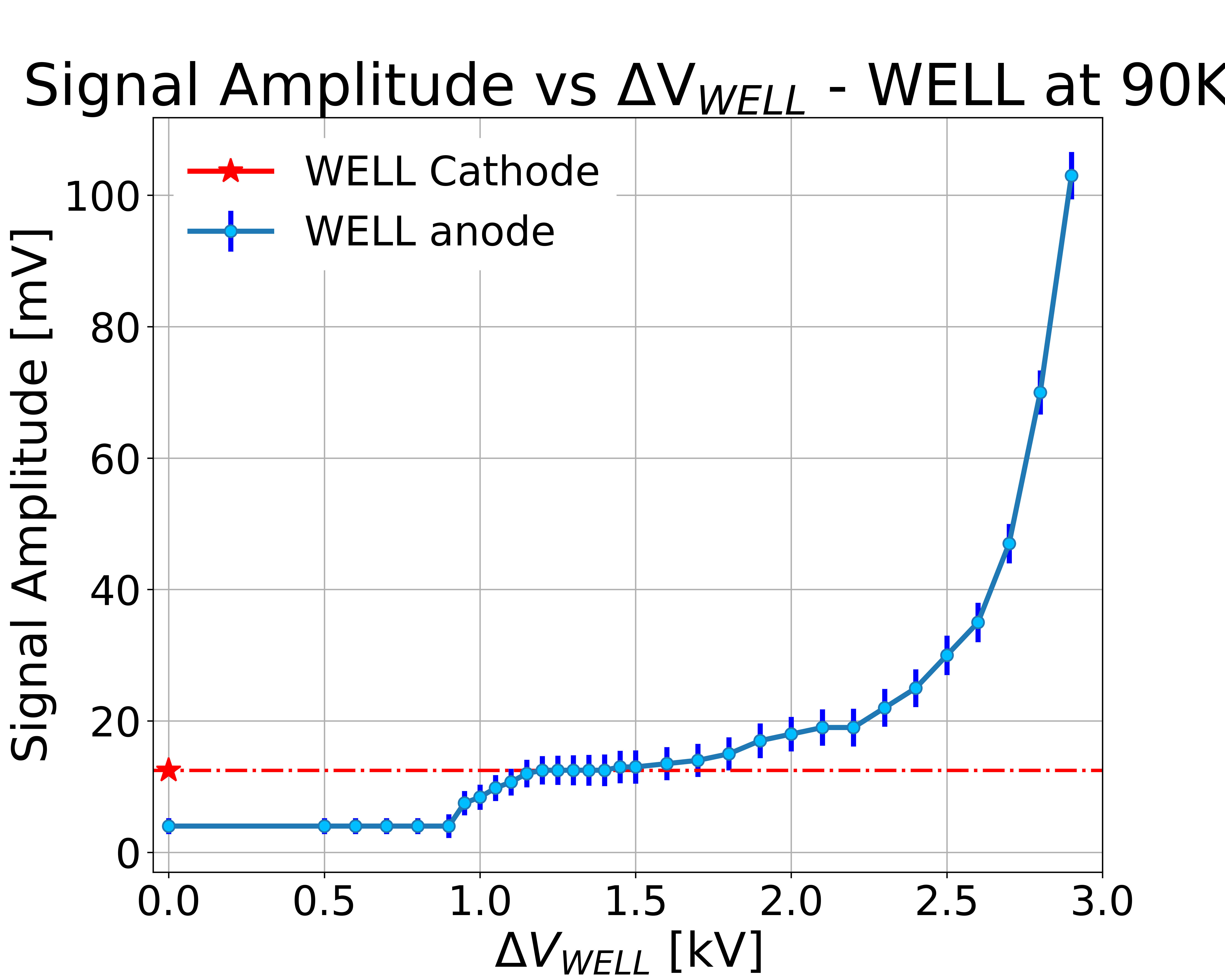}}
\hfill
\subfloat{
\includegraphics[width=7.5cm]{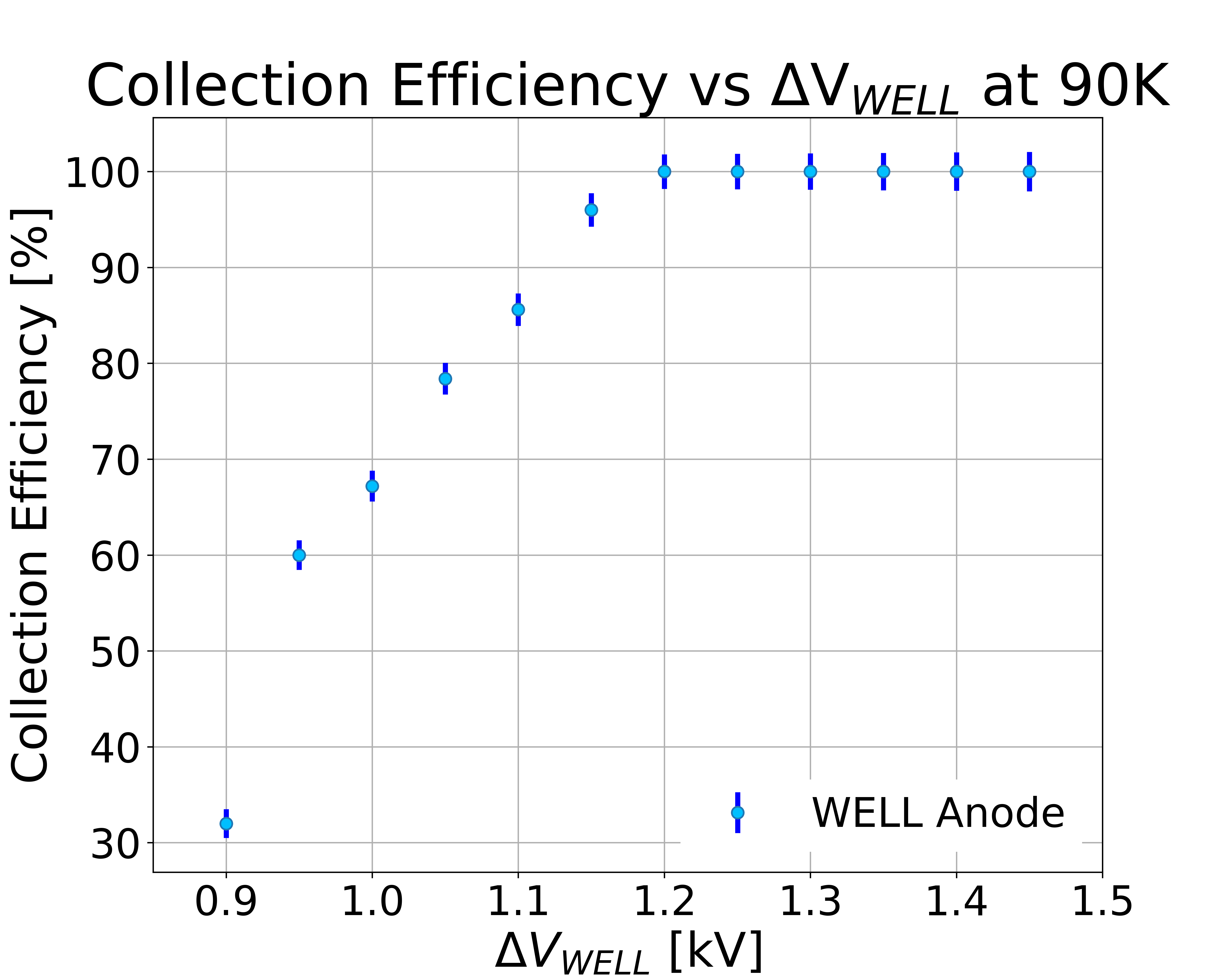}}
    \caption{\footnotesize \textbf{Left}: Scan of the WELL detector operated at 90~K, 1.2~bar in terms of the anode signals' amplitudes from the preamplifier as a function of $\Delta V\mathrm{_{WELL}}$, with E$\mathrm{_{d}}$ = 0.5 kV/cm. The cathode signal amplitude is also reported for comparison (see red star); \textbf{Right}: WELL collection efficiency as a function of $\Delta V\mathrm{_{WELL}}$, with E$\mathrm{_{d}}$ = 0.5 kV/cm.} 
    \label{fig:coll_eff}
\end{figure}

The process of primary charge drift and collection is independent of the adopted detector configuration.
An average collection signal from a dataset of 1000 waveforms is shown in Fig.~\ref{fig:CollectionSignals}, left. 
Its risetime provides a good estimate of the cathode-anode transit time of the ionization electrons, given that any alpha track orientation would induce signals of different shape yet with the same total duration. Thus all signals peak at about the same time when recorded through a charge amplifier. For our typical drift field of E$_d$ = 0.5~kV/cm, the average risetime (from 10\% to 90\% of the signal maximum) corresponds to 5.5~$\mu$s, yielding a drift velocity estimate of $v_e =$ 2.7~mm/$\mu$s, which is in approximate agreement with the simulated value at 90~K and 1.2~bar (from Magboltz \cite{BIAGI}).\\
\noindent
In Fig.~\ref{fig:CollectionSignals}, right, a typical spectrum of collection-signal amplitudes recorded for a sample of 4.5~$\times$~10$^3$ waveforms is shown. The spectrum is normalized to the count value of the most probable amplitude. It was calibrated by injecting a known amount of charge through a 2~pF capacitor. The mean value of the distribution corresponds to $\sim$1.5~x~10$^5$ primary electrons.\\
%{\color{red}{- Figure 5, left: I don't think this average signal tells the reader much, especially since there is a large variation in the angle of the alpha events. It only masks the noise - one may see on a single waveform - disappear, and one can not read of the rise-time with ease. I suggest replacing this plot by a histogram of the rise-time for every of the 1000 waveforms in the average. Alternatively, show how the geometry and the corresponding calculations with the Magboltz value create an average waveform like this.}}
\noindent
In Fig.~\ref{fig:coll_eff}, left, a scan of the WELL detector operated at 90 K, 1.2~bar was carried out in terms of the anode-signal amplitude from the preamplifier as a function of $\Delta V\mathrm{_{WELL}}$, with E$\mathrm{_{d}}$ = 0.5 kV/cm. The cathode-signal amplitude recorded at $\Delta V\mathrm{_{WELL}}$ = 0 kV and  E$\mathrm{_{d}}$ = 0.5 kV/cm, is also reported for comparison; it is the same for the RWELL (Fig.~\ref{fig:CollectionSignals}, left). Each point was recorded using the digital oscilloscope and represents the mean value of the distribution of signal amplitudes, each one containing 10$^3$ entries. The uncertainties were extrapolated as the $\sigma$ of the charge distributions. One can observe in Fig.~\ref{fig:coll_eff}, right, a rise of the collection efficiency reaching a plateau at about $\Delta V\mathrm{_{WELL}}$ = 1.2 kV, remaining constant until the onset of charge multiplication at about $\Delta V\mathrm{_{WELL}}$ = 1.5 kV. 
A charge gain $\sim$8 can be derived from Fig.~\ref{fig:coll_eff}, left, at $\Delta V\mathrm{_{WELL}}$ = 2.9 kV.

\noindent
Thus, for $\Delta V\mathrm{_{WELL}} \ge$  1.5 kV, the detector operates with full collection efficiency and, consequently, G$_\mathrm{Eff}$ (defined in Eqn.~\ref{eqn:gain}) is equivalent to the absolute detector gain G$_\mathrm{Abs}$. These conclusions apply to the RWELL case as well.

\subsection{20G$\Omega$-RWELL at 90~K - Amplification signals and spectra}
\begin{figure}[H]
\subfloat{
    \includegraphics[width=7.5cm]
    {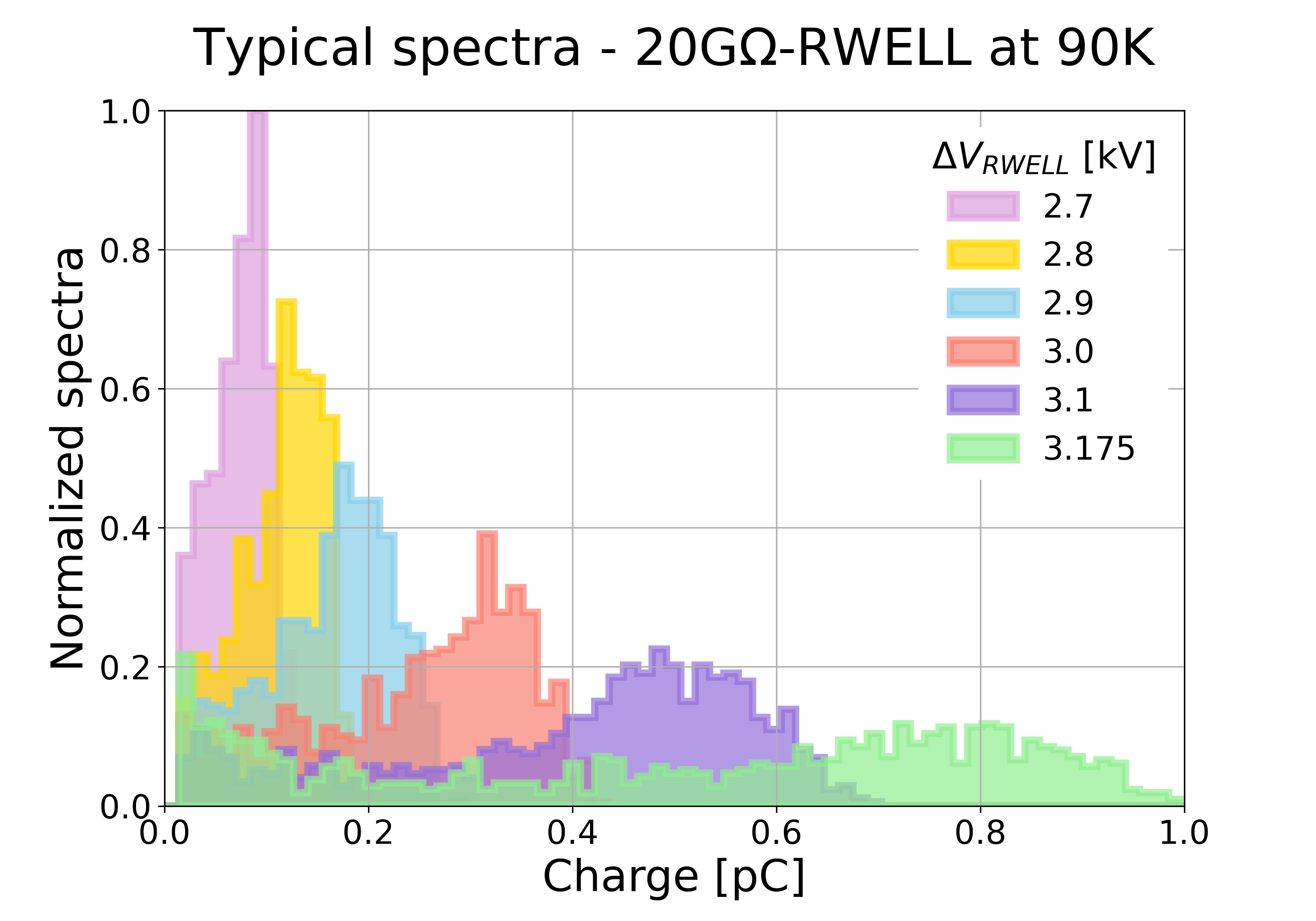}}
\hfill
\subfloat{
    \includegraphics[width=7.5cm]
    {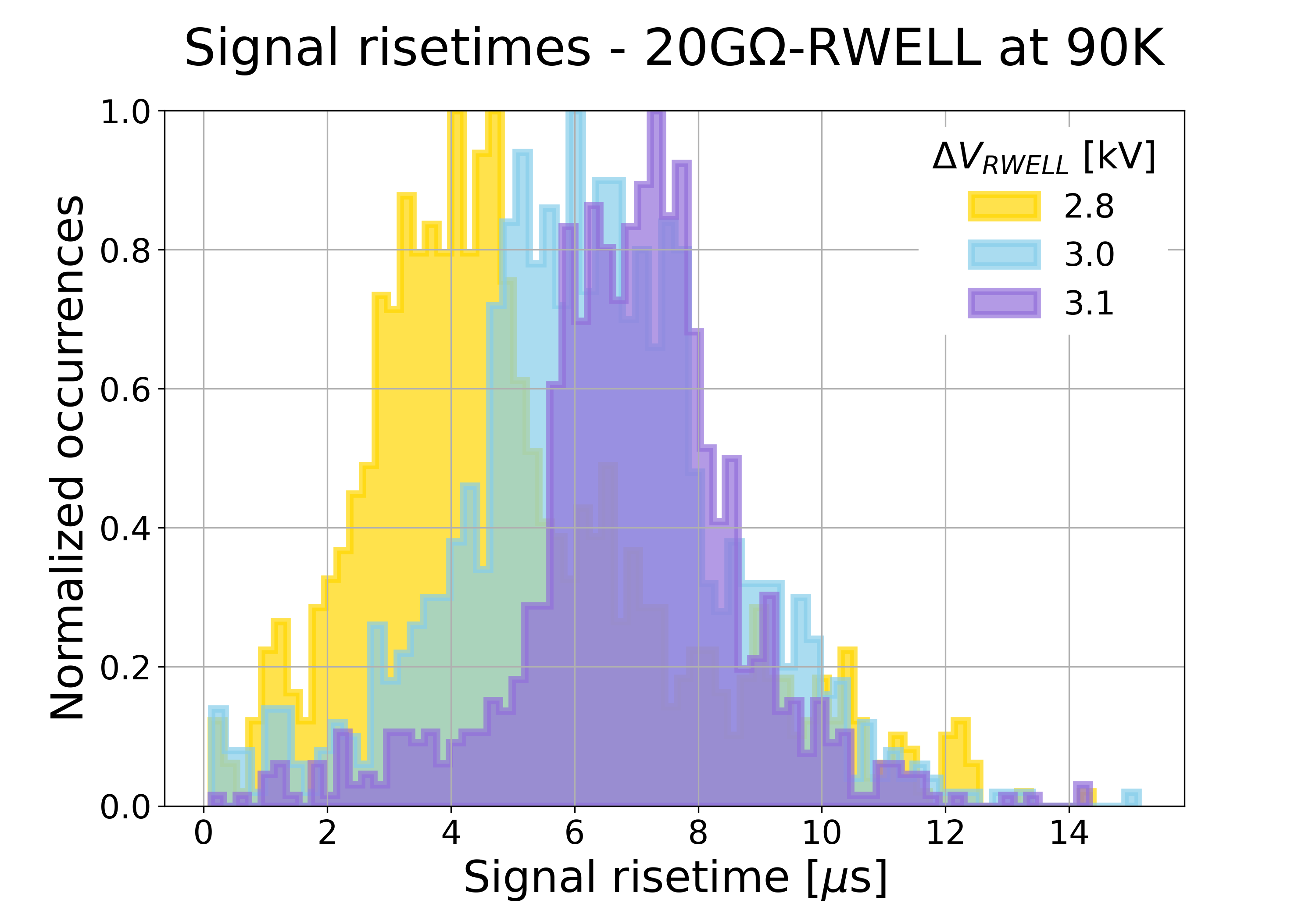}}
    \caption{\footnotesize \textbf{Left}: reconstructed $\alpha$-spectra obtained from an RWELL detector irradiated with $\alpha$-particles at 90~K, 1.2~bar; \textbf{Right}: risetime distributions obtained in the same thermodynamic conditions.} 
    \label{fig:Amplification}
\end{figure}

%In Fig.~\ref{fig:Amplification}, left, examples of average risetime distributions of $\sim$1000 normalized waveforms recorded from the RWELL anode are shown. 
The detector voltage was scanned in the range $\mathrm{\Delta V_{RWELL}}$~= 2.7-3.2~kV.% Waveforms were prepared for averaging by applying a smoothing moving-average algorithm.
 In Fig.~\ref{fig:Amplification} left, the calibrated charge spectra are shown. Histograms were recorded using the MCA and each one contained $\sim$5×10$^3$ entries. All spectra were normalized to their area and to the maximum at $\mathrm{\Delta V_{RWELL}}$ = 2.7~kV. The low-energy tail is attributed to the partial energy deposition of alpha particles exiting the collimator at large angles.

\begin{figure}[H]
    \centering
    \includegraphics[height=6cm,width=9cm]{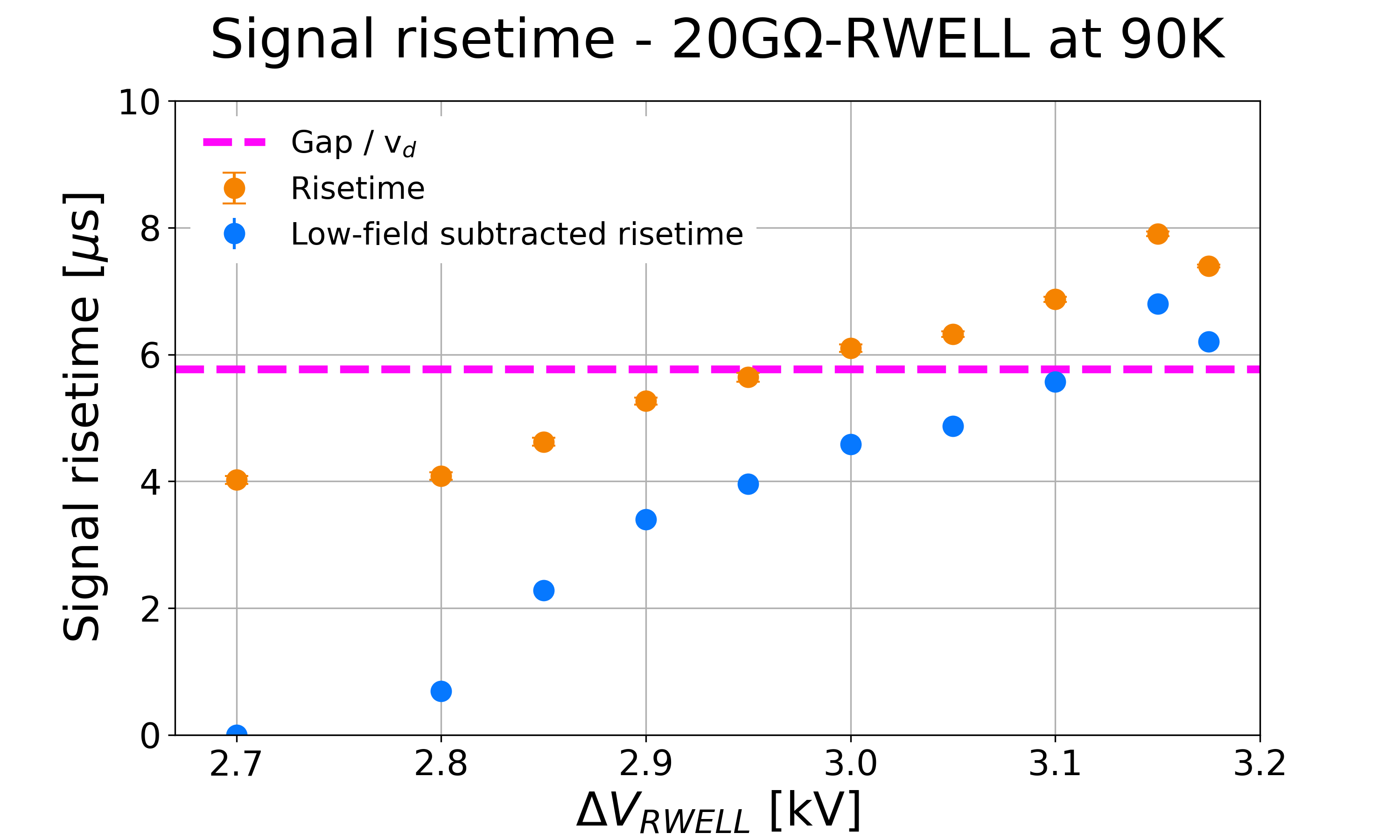}
    \centering
    \caption{\footnotesize Most probable signal risetime as a function of the applied voltage for the RWELL detector operated at 90~K, 1.2~bar, with E$_d$ = 0.5 kV/cm (orange circles). The low-subtracted risetime (blue circles) and the photon feedback time constant (magenta dashed line) are also reported.} 
    \label{fig:risetime}
\end{figure}

\noindent
For detailed risetime studies, a waveform-by-waveform analysis was made and the most probable value ($mpv$) was taken from a Gaussian fit to the risetime distributions.
Exemplary histograms are shown in Fig.~\ref{fig:Amplification} right. An increase of the risetime as a function of $\mathrm{\Delta V_{RWELL}}$ was observed in Fig.~\ref{fig:risetime} (orange circles), from around 4~$\mu$s to 8~$\mu$s. Despite the limited signal/noise, a drift velocity analysis was carried out using the technique introduced for $\alpha$-particles in \cite{Diana} and a constant value of $v_e \sim 1.8$ mm/$\mu$s was found (about 30$\%$ less than the simulated value quoted in the previous section). This indicates that the main contribution to the spread of the risetime distributions comes from the different orientations of the $\alpha$-particles, while the increase of the $mpv$ with $\mathrm{\Delta V_{RWELL}}$ stems from the detector response. A quadratic subtraction of the low-field risetimes, performed in order to isolate the effect, is superimposed in Fig.~\ref{fig:risetime} (blue circles). The time constant associated with photon feedback \cite{Fonte, Peskov} is shown for illustration (magenta dashed line).
Even if a good match is visible, the observed agreement might be coincidental. %photon feedback is driven by multiple cycles and 
Field modifications due to ion space charge or charge accumulation at the plate may induce a similar effect given that the contribution from the ion-drift across the THGEM structure amounts to $\approx$1 $\mu$s: a field reduction down to 20$\%$, even if a priori very extreme, would cause a similar increasing trend with $\mathrm{\Delta V_{RWELL}}$. On the other hand, ion feedback from the cathode, at the timescale of ms, would lead to well-separated spurious signals that were only present at $\mathrm{\Delta V_{RWELL}}$ = 3.175 kV (up to 40$\%$ of the events, with a discharge rate of $\approx1\%$). %In closing, the exact mechanisms leading to the breakdown above this value could not be unambiguously elucidated. 
Perhaps the best indication that breakdown is slow (either photon or ion-driven), in contrast to a conventional streamer mechanism, is the fact that the surface resistance required in present conditions is considerably larger than that required in VUV-quenched gases at room temperature.

\begin{figure}[H]
    \centering
    \includegraphics[width=11cm]{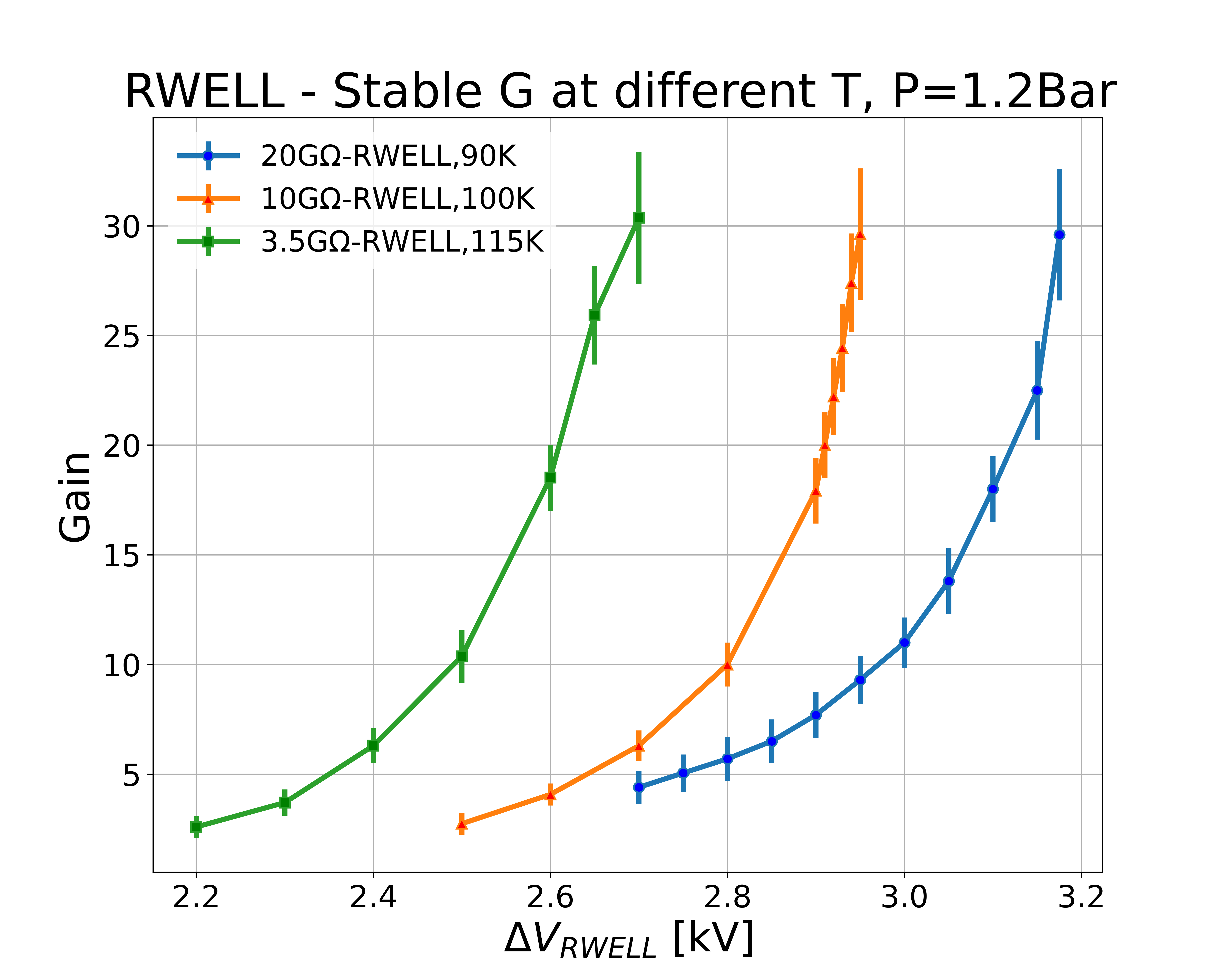}
    \centering
    \caption{\footnotesize RWELL gain, measured in stable operation conditions, as a function of the applied voltage at different temperatures. The resistivity of the DLC layer at each temperature is also reported. Reprinted from \cite{Tesi_2023}.}
    \label{fig:GainTemperatures}
\end{figure}

\noindent
Fig.~\ref{fig:GainTemperatures} depicts the RWELL gain as a function of $\Delta$V$\mathrm{_{RWELL}}$ at different temperatures (presented in Fig.~2a of \cite{Tesi_2023}). Gain uncertainties represent the gain variations of the stabilization curve. 

\noindent
Since the gas pressure was kept constant along the experiments, in first approximation the gas density is only temperature-dependent, $\rho_g\sim T^{-1}$. The DLC resistivity decreased with increasing temperature, from 20~G$\Omega/\square$ at 90~K to 10~G$\Omega/\square$ at 100~K and 3.5~G$\Omega/\square$ at 115~K \cite{Leardini2022}.
The detector characterization was performed first at 90~K, then the LAr level was reduced by warming up the system and extracting gas until the temperature reached the levels of 100~K and 115~K. 
The maximal achievable stable gain in the presence of quenched discharges was approximately 30 regardless of the temperature and DLC resistivity. 
The three curves manifest the same exponential trend, which is a feature of avalanche multiplication. Above a detector gain close to 18, the presence of quenched discharges was observed ($\approx$~0.25~$\mu$C); they did not prevent the RWELL operation and did not affect its performance. Above a gain of 30, the appearance of constant currents led to the detector tripping.

\subsection{Comparative study}
Past studies with LEM structures in dual-phase argon TPCs were carried out in different conditions: 87~K, 0.98~bar, cosmic muons, and unknown argon purity \cite{Wu_thesis}. 
Thus, such results in terms of maximum achievable gain can not be directly compared to the ones obtained in this study.
In order to demonstrate the potential advantages of the resistive configuration, the RWELL detector response was compared with that of THWELL and THGEM (LEM) in the same experimental system, irradiation source and conditions, at 90~K and 1.2~bar. Note that the RWELL was operated with DLC anodes of 200~M$\Omega/\square$ and 20~G$\Omega/\square$. The data were recorded after a gain stabilization cycle in purified Ar (see above). In all cases, $\mathrm{E_d}$ was kept at 0.5~kV/cm and the induction field in the THGEM configuration was set to 5~kV/cm. 
\begin{figure}[H]
    \centering
    \includegraphics[height=10cm,width=12cm]{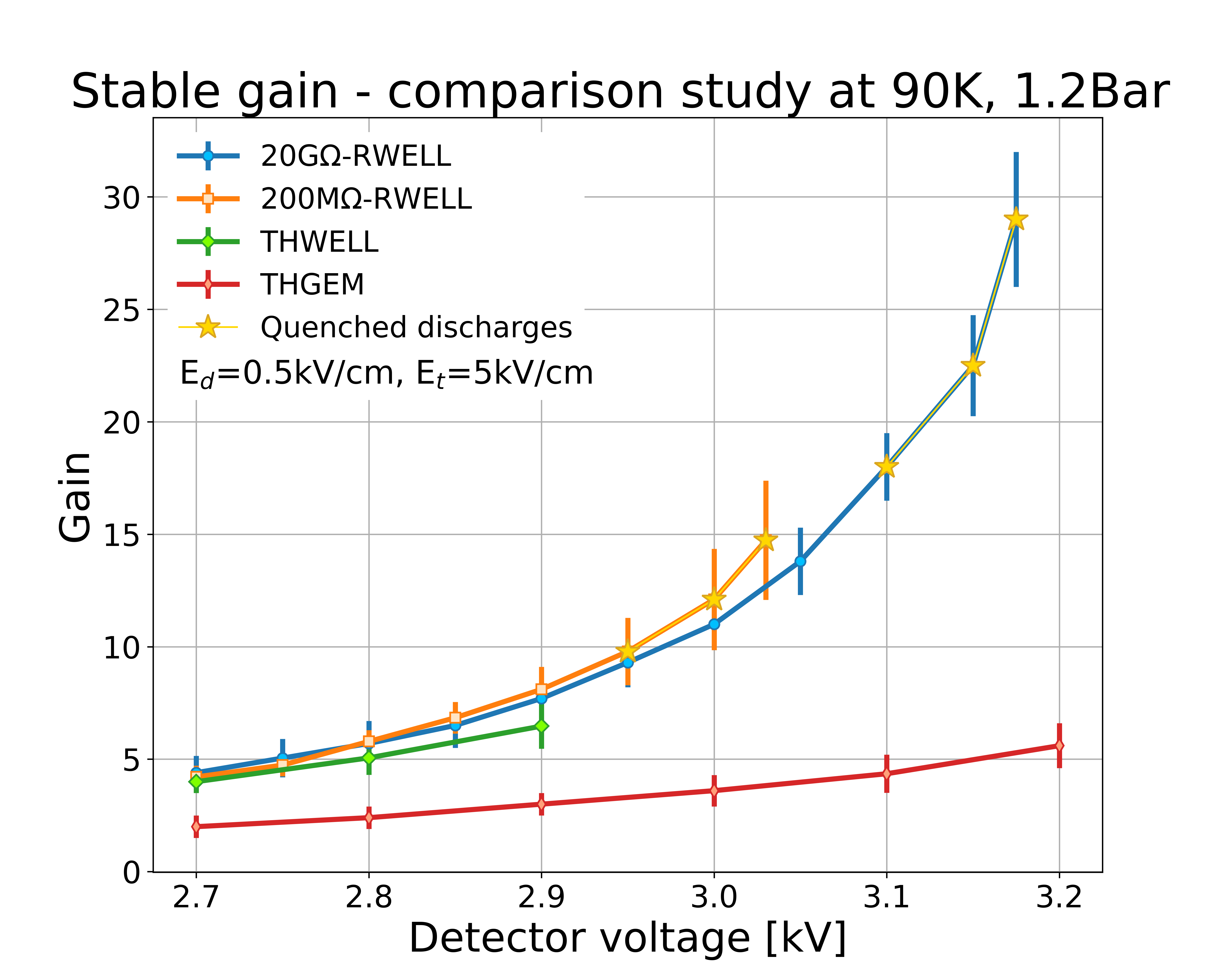}
    \centering
    \caption{\footnotesize Comparison study showing the stable gain for a THGEM+2mm induction gap, a THWELL, a 200M$\Omega$-RWELL and a 20G$\Omega$-RWELL, at 90~K and 1.2~bar.
    In all cases, the THGEM structure was 0.8~mm thick (see text).} 
    \label{fig:GainComparison90K}
\end{figure}
\noindent
Gain curves for all the tested detectors are shown in Fig.~\ref{fig:GainComparison90K}. The uncertainties on the gain were extracted from the gain stabilization measurements as the $\sigma$ of the distribution. It is possible to observe that both RWELL detectors outperform their non-protective counterparts.
The maximal stable-gain values achieved are summarized in Table~\ref{table_g}:

\begin{table}[H]
\begin{center}
\begin{tabular}{ |c|c|c|c|c| } 
\hline
Detector & THGEM & THWELL & 200M$\Omega$-RWELL & 20G$\Omega$-RWELL\\
\hline
Max stable Gain & 6 & 8 & 15 & 30\\ 
\hline
\end{tabular}
\end{center}
\caption{\footnotesize Maximal stable-gain values for the investigated structures.  }
\label{table_g}
\end{table}

\noindent
The respective 2- and 4-fold gain increase of the RWELL configuration compared to the THWELL can be explained by the discharge quenching features of the resistive layer. Discharges occurring in the two non-protected detectors, WELL and THGEM, generally led to the tripping of the detector and often damaged the preamplifier. In both RWELL configurations, the presence of quenched discharges was observed: at 10~<~G~<~15 for the 200M$\Omega$-RWELL and 18~<~G~<~30 for the 20G$\Omega$-RWELL. Discharges could be sustained throughout the detector operation without affecting the stable gain, the discharge charge ($\approx$0.25~$\mu$C) was $\sim$15-fold lower than the one of unprotected detectors ($\ge$~3.75~$\mu$C), and destructive effects to the electronics were strongly mitigated. Beyond $\Delta$V$\mathrm{_{RWELL}}$~=~3.05~kV for the 200M$\Omega$-RWELL and $\Delta$V$\mathrm{_{RWELL}}$~= 3.175~kV for the 20G$\Omega$-RWELL, both RWELL configurations were not operable due to the presence of constant currents. \\

\subsection{Discharge probability}
\label{section:discharges}
We define the discharge probability P$\mathrm{_d}$ as the number of discharges occurring per detected alpha particle per unit of time: 
\begin{equation*}
\mathrm{P_d = \frac{N_d}{E_r\times t}}
 \end{equation*}
\noindent
where N$\mathrm{_d}$ represents the number of discharges recorded, E$\mathrm{_r}$ is the rate of detected events and t is the measurement time. E$\mathrm{_r}$ was measured using a surface-barrier silicon detector\footnote{Ortec F Series Partially Depleted Silicon Surface Barrier Radiation Detector} and was found to be $\approx$10 Hz.\\

\begin{figure}[H]
    \centering
    \includegraphics[width=11cm]{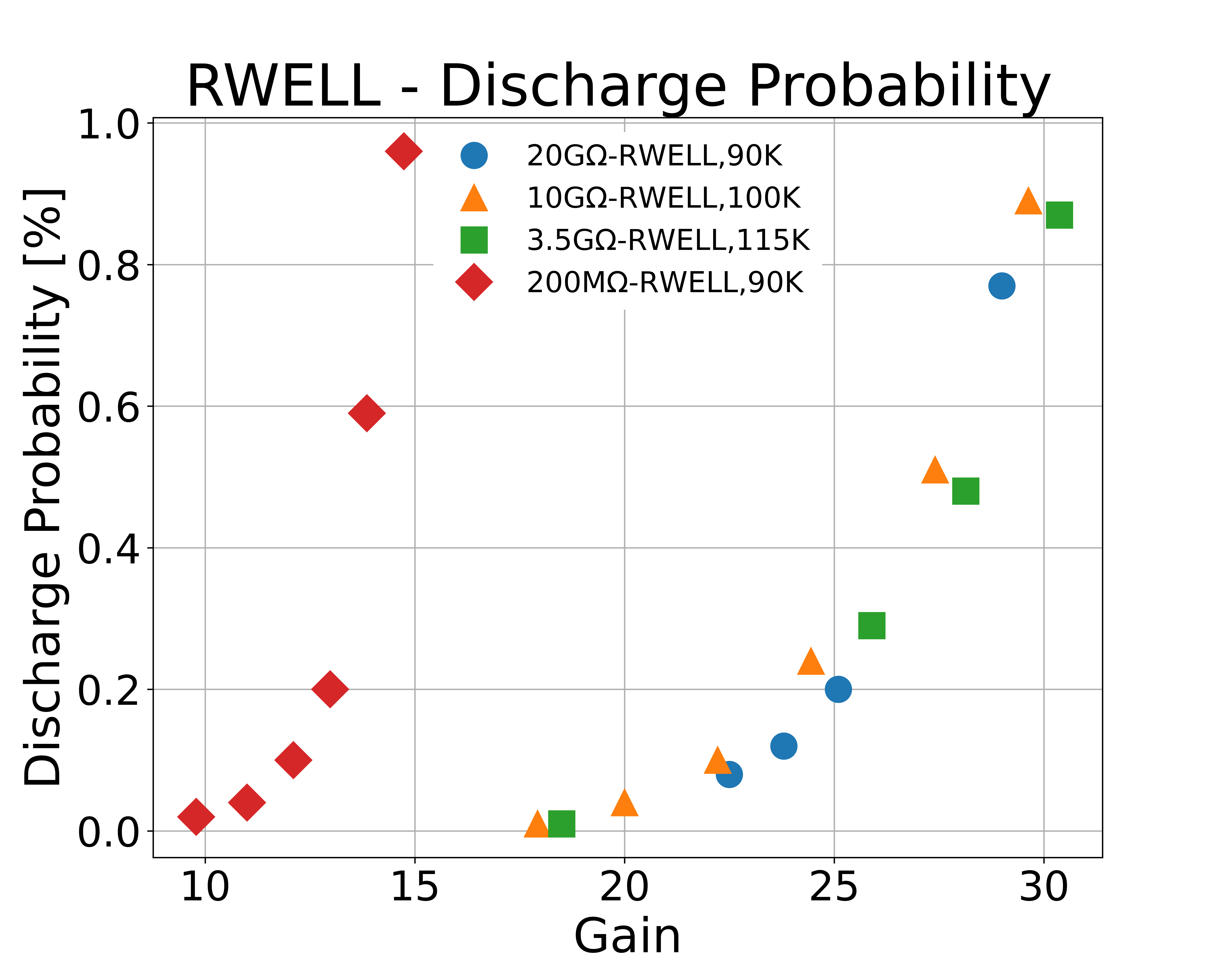}
    \centering
    \caption{\footnotesize Discharge probability as a function of the RWELL gain for four different configurations: 200M$\Omega/\square$ at 90~K and 20~G$\Omega/\square$ at 90~K, 10~G$\Omega/\square$ at 100~K, 3.5~G$\Omega/\square$ at 115~K. Irradiation done with 4~MeV alpha particles, pressure: 1.2 bar.} 
    \label{fig:Pd_gain}
\end{figure}

\noindent
In Fig.~\ref{fig:Pd_gain}, the discharge probability of the RWELL is depicted as a function of gain at different temperatures (90~K - 200~M$\Omega/\square$, 90~K - 20~G$\Omega/\square$, 100~K - 10~G$\Omega/\square$, 115~K - 3.5~G$\Omega/\square$).

\noindent
Note the fast rise in discharge probability with the lowest resistivity RWELL, while the other three behave similarly. This is related to the fact that, for a higher R$\mathrm{_S}$, the charge evacuation process is slower and consequently, the effective electric field at the bottom of the hole is reduced during the development of the avalanche, leading to quenching.
In all cases, at maximal gain values, the discharge probability reaches $\sim$1$\%$. 

\section{Summary $\&$ Discussion}
\noindent
The availability of DLC films with tunable surface resistivity has permitted, for the first time, the implementation of the cryogenic RWELL. It was operated under 4~MeV alpha particles in purified Ar at 90~K and 1.2~bar. With resistive anodes of 200~M$\Omega$/$\square$ and 20G~$\Omega$/$\square$, the RWELL exhibited discharge-quenching capabilities and could reach gains of $\sim$15 and $\sim$30, respectively. Relative to RWELL detectors operated at room temperature, much higher R$_S$ values were needed to reach lower discharge quenching factors \cite{Arazi_2014, Jash:2022bxy}. 
A comparison with a THWELL and with a THGEM~(LEM) +~2mm induction gap demonstrated the superiority of the resistive configurations in terms of stability and gain (the latter was 6 and 8 for the THGEM and THWELL, respectively, with alpha particles).\\

\noindent
The RWELL was operated also in the presence of moderate discharges, the latter being $\sim$15-fold smaller than those in THWELL. 
In general, non-resistive configurations could not be operated in the presence of discharges, which also happened to damage the electronic readout.  For all the configurations with R$\mathrm{_S}$ > 3.5~G$\Omega/\square$, it was possible to operate the detector with P$\mathrm{_d} \le$ 0.01$\%$ (upper bound estimation limited by the source rate and measurement time) up to a gain $\sim$18, after which the discharge probability sharply increased to $\approx$1$\%$ at a gain of $\sim$30.
The dependence of the discharge probability on R$_S$ cannot be explained by simple considerations based on the streamer mechanism, which would lead to quenching resistances lower than the ones employed here (in line with observations in quenched mixtures at RT). For this reason, the process of discharge quenching at cryogenic conditions requires further investigation. \\
\noindent
 A factor 2.5-5 higher stable gain values was reached with the RWELL compared to a bare THGEM/THWELL in our current setup under the same conditions. This demonstrates the potential advantage of coupling a resistive anode to the multiplier, also in noble gas at cryogenic conditions. A direct comparison with the results presented in literature obtained with LEM detectors \cite{Cantini_2014} was not possible due to the different conditions of operation (e.g. gas density and purity, type of primary ionization, different electrode parameters, etc.).\\

\noindent
To conclude, the successful operation of the RWELL detector at cryogenic
temperature represents an important milestone. Though requiring further investigations, it has the potential for improved scalable readout in future large-volume dual-phase detectors. The scaling up depends on industrial THGEM-electrode production limits and DLC-coating facilities. Electrode sizes reaching 50x50 cm$^2$ were already reported \cite{Aimard}; DLC-film coating currently reaches 20×100 cm$^2$ \cite{Song_DLC}. DLC films proved to be robust, minimally affected by aging, and cheap to produce \cite{Lv_2020}.\\ 
\noindent

\acknowledgments
We would like to thank Prof. Zhou Yi from the University of Science and Technology of China for the production of the DLC layers, Mr. Y. Asher for the technical help with the mechanical productions, Mr. Y. Shahar and Dr. M. Rappaport for their support with the use of the vacuum technology and cryogenics, and Dr. Ryan Felkai for his assistance during the measurements.\\

\noindent
This work was supported by Sir Charles Clore Prize, by the Nella and Leon Benoziyo Center for High Energy Physics, and CERN-RD51 ‘common project’ fund, the Xunta de Galicia (Centro singular de investigación de Galicia, accreditation 2019- 2022), and the “María de Maeztu” Units of Excellence program MDM-2016-0692. DGD was supported by the Ramón y Cajal program (Spain) under contract number RYC-2015-18820.

\bibliographystyle{JHEP}
\bibliography{bibliography}
\end{document}